# The Flattened Dark Halo of Polar Ring Galaxy NGC 4650A:

# A Conspiracy of Shapes?

Penny D. Sackett and Hans-Walter Rix[1]
Institute for Advanced Study, Princeton, NJ 08540

Brian J. Jarvis[2]
European Southern Observatory, La Silla, CHILE

and

Kenneth C. Freeman
Mt. Stromlo Observatory, Canberra, AUSTRALIA

astro-ph/9406015    6 Jun 94

## ABSTRACT

Kinematics and photometry of the polar ring galaxy NGC 4650A, including new observations of the rotation and velocity dispersion of its central stellar disk, are used to infer the presence of a dark matter halo and to measure its shape. Fits to the observed disk and polar ring rotation curves from detailed mass and photometric modeling rule out a spherical dark halo. The best fit models have halos with isodensity surfaces that are flattened to a shape between E6 and E7 (axis ratios between 0.4 and 0.3); the asymptotic equatorial speeds of these models are in excellent agreement with the I-band Tully-Fisher relation. This degree of dark halo flattening is larger than that expected from N-body collapse simulations of dissipationless dark matter. Since the kinematics and surface brightness profile of the central luminous body indicate that its light has an intrinsic axis ratio $c/a \lesssim 0.4$, in NGC 4650A the radial "conspiracy" between the dark and luminous components that leads to flat rotation curves may extend to the *shape* of the mass distribution as well.

*Subject headings:* dark matter — galaxies: halos — galaxies: kinematics and dynamics — galaxies: individual (NGC 4650A) — galaxies: structure — polar ring galaxies

## 1. Introduction

Dark matter halos have generally been assumed to be spherical, despite the lack of strong observational evidence to support this hypothesis. Although theoretical arguments based on the stability of disk galaxies against bar formation have been used in support of dark halos over dark disks (Ostriker & Peebles 1973), it is now believed that hot disks (Athanassoula & Sellwood 1987) and bulges (Kalnajs 1987) may be more efficient stabilizers than extended spherical halos. The HI rotation curves of normal spiral galaxies that have indicated (together with the assumption of Newtonian gravity) the presence of large amounts of radially extended unseen matter, are unable to place constraints on the *vertical* distribution of this dark matter because kinematics are usually measured only in the plane of the stellar disk. In order to study the flattening of the dark matter distribution, a luminous tracer must be found that probes the potential far from the galactic plane. Such studies have been carried out for only a handful of systems to date; the results are mixed.

The axis ratio of the massive dark halo of the Milky Way has been estimated by Binney, May & Ostriker (1987), assuming that the kinematics of extreme Population II stars are similar to those of the dark constituents in the halo. Using Stäckel potentials as a mass model for the Galaxy, this study concluded that the mass distribution of the halo must have a vertical-to-radial axis ratio between 0.3 and 0.6 (or between E7 and E4 in the notation used to describe the shape of elliptical galaxies) in order to match the observed anisotropy of the velocity dispersion characteristic of the extreme Population II stars.

By measuring the flattening of the extended X-ray isophotes around the elliptical galaxy NGC 720, Buote and Canizares (1994) have used the assumption of hydrostatic equilibrium to derive the shape of the underlying mass distribution of this system. They find that the axis ratio of the total gravitating matter lies between E5 and E7 that is, slightly *flatter* than the E4 shape of the projected light. After estimating the mass distributions of the stars and gas from surface brightness profiles, they fit the dark matter directly and find shapes in good agreement with those derived for the total mass in the system.

The flattening of the total mass distribution of the S0 galaxy NGC 4753 has been estimated by Steiman-Cameron, Kormendy & Durisen (1992) to lie between E0.1 and E1.6 on the basis of fitting an inclined, precessing disk model to the dust lanes seen against the stellar light, which has a projected E4 shape. The goodness-of-fit of their excellent model for the complicated dust patterns is independent of the flattening of the underlying potential; the flattening determination is based on the assumption that the gas orbits are filled and have completed at least 6 rotation periods at all radii. With these assumptions, mass distributions flatter than E1.6 would have caused the dusty disk to have precessed more than is observed. If the total mass in this system is considerably *rounder* than that of the light, as their work would indicate, the dark matter in NGC 4753 would dominate the potential inside the optical radius, contrary to what is found in most early spirals.

Polar ring galaxies (PRG) are early-type galaxies surrounded by outer rings of gas, dust and stars on orbits that are nearly perpendicular to the plane of the flattened central galaxy, which rotates about its own apparent minor axis (for reviews, see Whitmore *et al.* 1990, Casertano, Sackett & Briggs 1991). Steiman-Cameron & Durisen (1982) pointed out a decade ago that PRG could be used as probes of the shape of galactic gravitational potentials. Polar orbits, unlike orbits in the equatorial plane, are quite sensitive to the vertical distribution of mass, and thus to the flattening of the dark matter halo. In an oblate flattened potential, closed polar orbits are elongated along the polar axis and have smaller rotational speeds as they cross the polar axis than do closed orbits in the equatorial plane at the same galactocentric radius. Schweizer, Whitmore & Rubin (1983) and Whitmore, McElroy & Schweizer (1987, hereafter WMS) were the first to apply this test; they compared the ratio of ring and disk rotation speeds at a galactocentric radius of $0.6\,R_{25}$ (*i.e.*, at about one-half the optical radius of the central galaxy) in each of three PRG to that which would be expected for orbits in a scale-free logarithmic potential. WMS concluded that the axis ratios of the potentials ranged from $0.86 \pm 0.21$ to $1.05 \pm 0.17$ in these three galaxies. Since for the logarithmic potential used by WMS the isodensity surfaces are roughly three times more flattened than the corresponding equipotentials (*cf.* Binney & Tremaine 1987, p48), the WMS results encompass mass distributions with *a wide range of flattenings*. The only polar ring galaxy known to have an elliptical (rather than an S0) as its central host, AM 2020-504, has been modeled by Arnaboldi *et al.* (1993), who found that a halo flattened by the same amount as the projected light (E4) produced a somewhat better fit to the ring kinematics than a spherical halo (E0).

The shape of dark matter halos is thus an open question. It is also an important one: a strongly flattened dark matter distribution would imply that the dark constituents are (or were) dissipative, and therefore probably baryonic. Recent results from experiments (Alcock *et al.* 1993, Aubourg *et al.* 1993, Udalski *et al.* 1993) searching for microlensing of stars in the Magellanic Clouds due to compact objects in our own halo (*i.e.*, MACHOs) lend credence to the idea that a measurable fraction of the dark halo of the Milky Way may be composed of baryonic matter. These results must be interpreted with care, since a substantially flattened halo would have a smaller optical depth to gravitational lensing toward the Large and Small Magellanic Clouds (Sackett & Gould 1993, Gould, Miralda-Escudé & Bahcall 1994), thus reducing the number of events that the microlensing experiments could detect for a given MACHO mass distribution.

One of the three systems studied by WMS was the prototypical polar ring galaxy NGC 4650A, for which these authors estimated that the vertical-to-radial axis ratio of the *isopotential* surfaces was $0.86 \pm 0.21$, implying an axis ratio of about 0.6 (E4) for the isodensity surfaces of the galaxy interior to $0.6\,R_{25}$. A more recent analysis of NGC 4650A (Sackett & Sparke 1990, hereafter SS) used the WMS dataset together with kinematics from other sources, and compared the entire extent of both ring and disk rotation curves to detailed mass models that separated the effects of the disk, ring, and halo potentials. They concluded that a range of dark halo flattenings from E0 to E8 (*i.e.*, axis ratios of the halo isodensity surfaces between 1.0 and 0.2) were permitted by the data, with an E6 dark matter halo giving the best fit. Since polar rings are nearly always gas-rich (Shane 1980, Schechter *et al.* 1984, van Gorkom, Schechter & Kristian 1987, hereafter GSK, Richter, Sackett & Sparke 1994), their kinematics are generally determined through emission-line spectroscopy in the Hα line of ionized hydrogen or the 21cm HI line of neutral hydrogen. The central bodies of polar ring galaxies, on the other hand, are nearly devoid of gas, hence their kinematics must be derived from difficult stellar absorption line measurements. Uncertainty in the kinematics of the central stellar body prevented SS from making a tighter determination of the shape of the dark matter halo for NGC 4650A.

The primary advantage of NGC 4650A over other systems for studies of the flattening of the dark halo is that the polar ring in NGC 4650A extends to much larger heights above the disk plane (at least 16 kpc or 20 disk scale lengths!) so that the effects of the luminous and dark matter can be cleanly separated. Furthermore, the wealth of data now available for this system allow us to build self-consistent models for its kinematics that are highly constrained.

This paper presents new observations of the kinematics of the central galaxy of NGC 4650A, and uses these new data together with other observational constraints to discriminate between mass models of differing dark halo oblateness, placing tight limits on the shape of the dark matter distribution of this galaxy. In §2, we present and discuss the new observations: improved absorption-line kinematics of the central disk and I-band photometry of the system. In §3, each component of the mass model we use for NGC 4650A is discussed together with the observations that constrain its free parameters. In this section, we also give analytic expressions in cylindrical coordinates for the gravitational force due to an ellipsoidal, pseudo-isothermal halo of arbitrary flattening. In §4, we describe how model orbits are calculated and analyzed in order to allow a direct comparison with the data. In particular, we discuss how we enforce radial hydrostatic equilibrium and model the effects of line-of-sight integration, seeing convolution, projection. The results of fitting the models to the observations are presented in §5; we show that observational constraints indicate the presence of a dark halo in NGC 4650A that is flattened, with isodensity surfaces of axis ratio $0.3 \lesssim c/a \lesssim 0.4$. Implications of this result are discussed in §6 where our findings are summarized.

## 2. Observations of NGC 4650A

### 2.1. New Kinematics of the Central Galaxy

The central stellar body of NGC 4650A was first shown to be rotating about its apparent minor axis by Schechter, Ulrich & Boksenberg (1984), which indicated that the system was an oblate stellar system encircled by a nearly polar ring, not a prolate one surrounded by an equatorial ring. Their result was confirmed by WMS, who found that the amplitude of the rotation was comparable to the central velocity dispersion. The substantial rotation, together with the exponential profile of the disk surface brightness, led WMS to the conclusion that the central stellar body was an S0 disk.

The low surface brightness of the stellar disk, compared to the night sky, makes the measurement of stellar kinematics more difficult than deriving the ring kinematics from emission line ring gas; this is especially true at large galactocentric radius. It was the uncertainty in the WMS disk rotation curve that prevented SS from making a tight determination of the shape of the gravitational potential of NGC 4650A. In addition, a sharp dip ($\sim$30 km s$^{-1}$) in the measured disk speeds at 2-3 scale lengths (12"-16") observed by WMS on both sides of the central galaxy was troubling; if real and due to steady-state gravitational effects, it would indicate a sharp dip in the surface mass density of the disk with no corresponding dip in luminosity. For these reasons, we undertook deep long-slit spectroscopy along the major axis of the central stellar component of NGC 4650A in order to obtain a more well-determined stellar disk rotation curve, and, if possible, to extend it to larger radius.

The long-slit data were obtained on 1 May 1990 using the Boller and Chivens spectrograph attached to the 3.6m telescope operated by the European Southern Observatory at La Silla, Chile. The detector was an RCA SID 503 CCD (1024 x 640 pixels) with a read-out noise of approximately 24 electrons. On-chip binning along the slit produced a pixel size of 1.65", and gave a final frame format of 210 x 1024 pixels. The grating was used in second order, yielding a dispersion of 59 Å/mm centered at $\sim$ 5500Å and a spectral range that included both the Mg b and Na I absorption lines. Table 1 lists the instrumental parameters. In average seeing of 2.2", a total of four 90-minute exposures were taken with the slit aligned along the major axis of the central galaxy, passing through the nucleus at a position angle of 241°. Exposures were interlaced with HeAr arc exposures to monitor the geometric stability of the instrument. Suitable standard template stars were also observed against which the rotation velocities and velocity dispersions could be measured.

Table 1: Instrumental Parameters

| | |
|---|---|
| Detector | RCA Hi-Res ESO CCD #8 |
| Spectrograph | Boller and Chivens |
| Frame format | 210 x 1024 (3 x 1 binning) |
| Dispersion | 0.872 Å/pixel |
| | 47 km s$^{-1}$/pixel |
| Pixel size | 1.65" x 0.872 Å (15$\mu$m pixels) |
| Wavelength range | 5085 – 5980 Å |
| Instrumental Dispersion | $\sim$68 km s$^{-1}$ |
| Slit width | 3" |
| Slit length | 2.9' |

The removal of CCD bias, flat-fielding with quartz-lamp and sky exposures, and cosmic-ray removal were performed using standard IRAF tasks. Transformation from pixel coordinates, $I = I(x, y)$, to wavelength–slit position coordinates, $I = I(\log \lambda, s)$, was achieved with the IRAF tasks *fitcoords* and *transform*, using the arc exposures and traces of the template stars. This transformation was applied to all galaxy and template star exposures, yielding images with the slit coordinate, $s$, in one direction and $\log \lambda$ in the other. The four separate major axis exposures were co-added into one image at this stage. At the galaxy center, where there is sufficient signal-to-noise in the individual constituent spectra, the velocity dispersion, $\sigma$, measured from the individual exposures agreed with that measured from the combined data.

Compensation for differences in the observed continuum shapes between the galaxy and template spectra was made by subtracting a fifth order polynomial from both. The velocities and dispersions were measured by shifting and broadening template spectra by trial amounts and comparing them to the galaxy data *in pixel space*; a Gaussian broadening function, parameterized by its mean, $v$, and dispersion, $\sigma$, was assumed. For each point along the major axis, the kinematic parameters $v$ and $\sigma$ are taken to be those that minimize the $\chi^2$ difference between the galaxy and template spectra. Errors in these parameters are defined by the $\Delta\chi^2 = 1$ contour in parameter space, which corresponds to a 68% (1-sigma) confidence limit. (The technique is discussed in Rix and White 1992 and in Rix *et al.* 1994, see p. 29.)

The errors calculated from $\chi^2$ do not reflect systematic errors, such as those due to imperfect sky subtraction or the use of a mismatched template spectrum. We estimated the effect of such errors by comparing the four individual galaxy frames to one another and analyzing the data using three different template spectra. Since the velocity dispersion of the galaxy is not much larger than the "instrumental dispersion," it is more susceptible than $v$ to systematic errors. Our derived values for $\sigma$ differ by $\sim$5 km s$^{-1}$ between these various methods of reducing the data; we take this as an estimate of the systematic errors in the reduction process.

Since the mass models can reproduce only the anti-symmetric part of the rotation curve and ring dust may affect measurements at intermediate radii, we have adopted a systemic radial velocity that produces an anti-symmetric rotation curve at the galaxy center and at large galactocentric radius. Our final disk velocities and velocity dispersions are plotted in Fig. 1 and listed in Table 2. This table also contains estimates of the statistical variances due to known noise sources (*i.e.*, photon noise and read-out noise).

Due to the higher S/N of our observations, our measured disk velocities have uncertainties that are about a factor of two smaller than those measured by WMS in the inner region of the disk. We do not see the 30 km s$^{-1}$ "dip" in rotation speeds at 12-16$''$ reported by WMS; within our smaller uncertainties, our measured disk speeds rise smoothly and linearly from the center to 12$''$, where they turn over and flatten. On the approaching (northeastern) side of the disk, a foreground star at $\sim$ 18$''$ along the major axis contaminated the spectrum; we have discarded spectra within $\sim$1.5$''$ of the star. It is possible that the spectra within 2.5$''$ may also contain a fraction of light from this foreground star, but we estimate that it would constitute no more than 15% of total signal. The velocity measurement at 3″ on the southwest side of the galaxy may be affected by foreground dust from the ring; the same effect is evident in the WMS data. The low disk surface brightness of NGC 4650A forced us to combine spectra at 20–30″ into one spatial bin in order to ensure sufficient S/N to determine a reliable line center. Within the large uncertainties at these radii, our velocities are consistent with the WMS speeds of 100–110 km s$^{-1}$, though it is unclear to us how their measurement was made since we estimate that their instrumental parameters yield S/N $\lesssim$1.0 per spectral resolution element at these distances.

Table 2: Disk Major Axis Kinematics for NGC 4650A

| R[″] | $V$[ km s$^{-1}$] | $\Delta V$ | $\sigma$[ km s$^{-1}$] | $\Delta\sigma$ |
|---:|---:|---:|---:|---:|
| -23.93 | -135.3 | 27.3 | 61.4 | 34.2 |
| -15.68 | -47.0 | 18.6 | 38.0 | 23.6 |
| -14.03 | -57.7 | 16.9 | 89.9 | 19.5 |
| -12.38 | -72.2 | 14.3 | 63.7 | 14.0 |
| -10.73 | -68.7 | 7.8 | 45.4 | 9.9 |
| -9.08 | -59.4 | 9.5 | 77.5 | 9.6 |
| -7.43 | -50.5 | 6.1 | 45.2 | 6.8 |
| -5.78 | -23.2 | 4.8 | 60.9 | 4.8 |
| -4.13 | -32.6 | 4.3 | 63.0 | 4.5 |
| -2.48 | -25.6 | 3.0 | 54.3 | 3.4 |
| -0.83 | -0.4 | 2.6 | 60.4 | 2.4 |
| 0.83 | 3.2 | 2.6 | 59.5 | 2.1 |
| 2.48 | 12.2 | 3.9 | 57.0 | 4.1 |
| 4.13 | 43.8 | 7.4 | 64.8 | 9.2 |
| 5.78 | 44.2 | 7.4 | 70.4 | 8.9 |
| 7.43 | 63.0 | 6.5 | 56.5 | 7.2 |
| 9.08 | 61.5 | 8.7 | 70.8 | 8.9 |
| 10.73 | 79.7 | 7.4 | 45.3 | 9.9 |
| 12.38 | 87.2 | 12.1 | 71.6 | 16.4 |
| 14.03 | 71.7 | 15.6 | 65.4 | 17.1 |
| 15.68 | 89.2 | 14.3 | 43.8 | 23.6 |
| 17.33 | 90.9 | 2.6 | 6.4 | 25.7 |
| 25.58 | 91.7 | 35.1 | 10.8 | 133.2 |

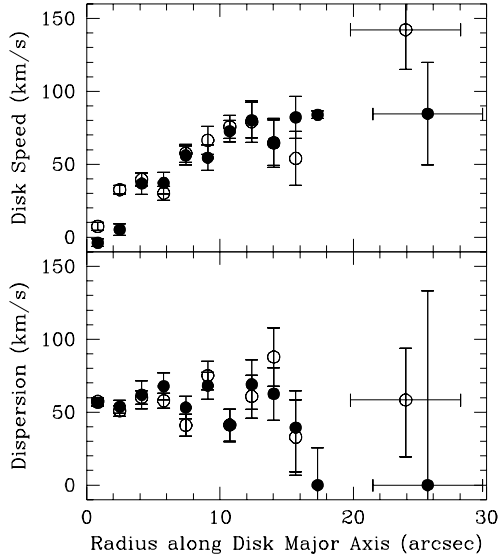

**Fig. 1.—** The observed stellar line-of-sight velocity (*top*) and velocity dispersion (*bottom*) as a function of galactocentric radius along the major axis of the central disk component of NGC 4650A. Velocities are derived from the new deep long-slit observations described in §2. Approaching (*eastern*) velocities are indicated by open circles, receding (*western*) velocities by solid circles. Error bars are $1\sigma$. The spectra near 18″ on the northeastern side of the galaxy were contaminated by a foreground star that lies along the major axis of the disk.

Our line-of-sight dispersions are symmetric about the photometric center of the galaxy, have a measured central value of about 60 km s$^{-1}$, and remain constant or decline very slowly to the last reliably-measured point at a radius of about 14″ (*i.e.*, 3 disk scale lengths). The dispersions derived by WMS are asymmetric with respect to the galaxy center, have much larger uncertainties, and decline with radius from about 90 km s$^{-1}$ at the center to about 50 km s$^{-1}$ at 10″. If the difference between their values on the two sides of the galaxy is taken as a measure of the true uncertainties, our measurements are consistent with WMS. Our higher spectral resolution and higher S/N give much tighter constraints on $\sigma(R)$. Since the dispersion measurements become much less reliable at large galactocentric radius, we experimented with templates of *fixed* dispersion (60 km s$^{-1}$) in order to derive the line-of-sight velocities in the disk. Such fixed-$\sigma$ fits were consistent with the results from free-$\sigma$ fits.

In order to check the extent to which the approximate constancy of $\sigma(R)$ with radius could be due to insufficient spectral resolution, we derived dispersions for unbroadened stars (with high S/N) using other stars taken during the same night as templates. The resulting dispersion due to pixel interpolation during the wavelength calibration and template mismatch was only about 0.4 pixels, or 19 km s$^{-1}$. When subtracted in quadrature, this would imply a ∼5% downward correction to $\sigma$ is indicated for the highest signal-to-noise (S/N > 13) galaxy spectra. Extending these tests into the low signal-to-noise regime (S/N=6.5) by adding gaussian noise to the stellar test spectra produced "spurious" dispersions rarely larger than about 0.52 pixels or 24 km s$^{-1}$—corresponding to a $\sim$8% downward correction from 60 km s$^{-1}$. Therefore, to correct for pixel interpolation and template mismatch, we have subtracted 19 km s$^{-1}$ in quadrature from $\sigma$ to obtain the dispersion measurements presented in Table 2, Fig. 1 and all subsequent plots.

## 2.2. New I-band Photometry

The I-band images of NGC 4650A were acquired on 30 March 1992 at the Mt. Stromlo and Siding Springs Observatory (MSSSO) 2.3-m telescope with the Nasmyth focal reducer, using a thinned Tektronix 1024 CCD and a filter that gives an accurate approximation to the Cousins I-band. The image scale was 0.60 arcsec pixel$^{-1}$. Weather conditions were photometric. E-region standards were observed several times during the night in order to transform the images onto the standard Cousins I-band system. The image in Fig. 2 was obtained by co-adding three 300-sec exposures and one 100-sec exposure; the shorter exposure was needed because the photon counts from the innermost few pixels of NGC 4650A in the 300-sec exposures approached the A/D saturation level of the CCD system. Conventional techniques were used to debias, flatfield and combine the images. The sky surface brightness was measured to be 18.84 I mag arcsec$^{-2}$. Faint residual fringing is visible in the image at a level below about 1% of sky. After masking out all foreground stars, we estimate that the total I-band magnitude of NGC 4650A is $m_I = 11.43 \pm 0.03$. By modeling the central galaxy as an inclined exponential disk (but giving no weight to regions with strong dust contamination) and then subtracting its light from the image, we estimate that the ring contributes $28^{+8}_{-6}$% of the total light, making the central disk about 2.5 times brighter than the ring in the I-band.

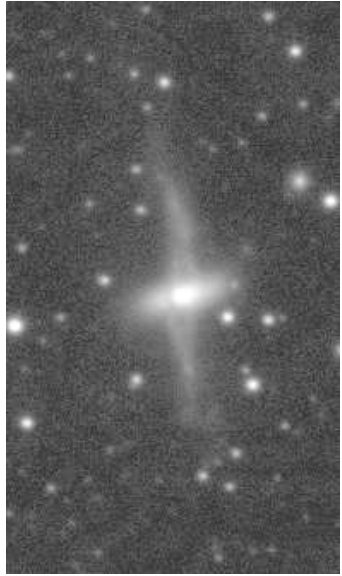

**Fig. 2.—** I-band image of NGC 4650A shown with logarithmic contrast. North and east are approximately up and to the right, respectively.

Fig. 3 shows the surface brightness profiles in the I-band along the major and minor axes of the central body of NGC 4650A. For comparison, the B-band profile from WMS along the major axis is also shown. The major axis profile is exponential, with a dip at about 5″ on the southwest side due to extinction from ring dust where the inner ring orbits pass in front of the disk; as expected, the dip is more pronounced in the B-band. In general, the shape of the profile is very similar in both bands. The color of the central disk of NGC 4650A is B-I = 1.8, and no significant large scale radial color gradient is evident in the central 15″. In the center of the galaxy, a narrow brightness enhancement comparable in width to the size of the 1.6″ seeing disk can be seen above the exponential profile in the B profile . This central feature appears to be bluer than the disk, although just how much bluer is difficult to estimate since the seeing was worse for the I-band (2.25″) photometric observations than for the B-band (1.6″). The bright feature at 18″ on the northwestern side of the galaxy is a red foreground star.

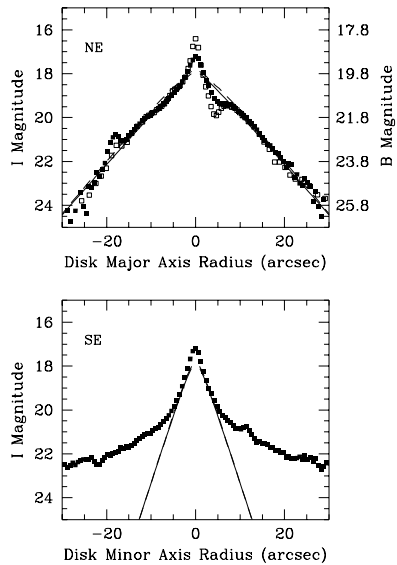

**Fig. 3.–** The surface brightness profile along the major (*top*) and minor (*bottom*) axes of the disk of NGC 4650A in the I-band (solid squares, this work) and in the B-band (open squares, from WMS). Note the offset in the magnitude scale for the two bands. The thin (solid) and thick (dashed) exponential disk models (§3.1) are shown superposed; the two fits are nearly identical, but imply different inclinations and scale lengths. Extinction at about 4″ can be seen on the western side of the major axis profile where dust in the polar ring passes in front of the disk. A star about 18″ along the eastern major axis of the disk is apparent as a local brightness peak.

Surface brightness profiles along the major axis of the ring in I (our image) and B (from WMS) are shown in Fig. 4. Light along the minor axis of the central object extends out to about 15″ in this profile. The ring is much bluer than the central body; at 30″ along the ring major axis the ring color is B-I ∼ 0.8. Bi-symmetric knots at about 30″ are seen in the B-band profile and in the I-band image (Fig. 2). The poorer seeing during the I-band observations degrades the sharpness of these peaks in the I-band profile. We interpret the knots as line-of-sight integration effects as the ring twists through edge-on at this radius, consistent with an increased brightness of about a factor of two.

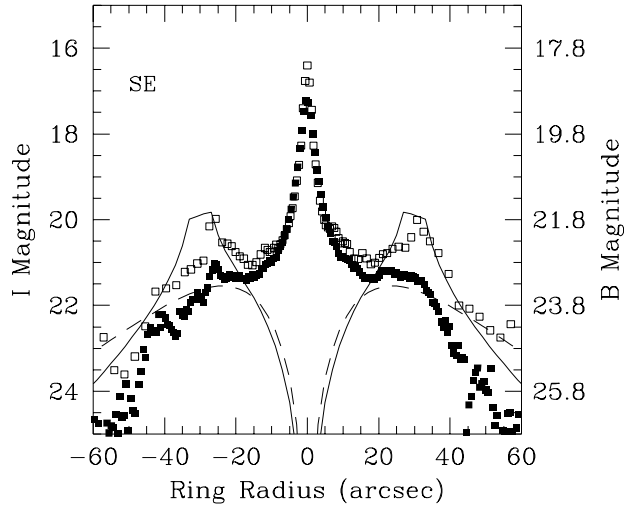

Fig. 4.– The surface brightness profile along the major axis of the polar ring of NGC 4650A in the I-band (solid squares, this work) and in the B-band (open squares, from WMS). A transparent ring model, warped according to the prescription described in §4.1, with mass $M = 2.4 \times 10^9 M_\odot$ and $M/L_B = 2.0$ is shown as the solid line; the dashed line indicates the same ring model with a constant inclination of 85°.

### 2.3. Ring Kinematics

The polar kinematic data set used here is identical to that used by SS, so only a brief review is presented here. Our analysis, however, and in particular our interpretation of the warp, is somewhat different.

Two different Hα observations are available for NGC 4650A: the long-slit data of WMS and the Fabry-Perot data of Nicholson (1989). In Fig. 5, the Hα line-of-sight ring velocities are shown as deduced by Nicholson (1989) from his modeling of the Fabry-Perot data and as measured by WMS along the major axis of the ring after our correction for the ring warp (see §4.1). The rotation curves derived from each of these two groups agree within the scatter. Neutral hydrogen synthesis mapping (GSK) in the 21-cm line at the Very Large Array indicates that the HI in NGC 4650A is associated with the polar ring, and that its kinematics are consistent with that derived from Hα studies wherever Hα and HI are spatially coincident. The HI ring extends about twice as far as the optical ring, but since the synthesized beam width in the B/C array is about 20″ (FWHP), the HI measurements have much poorer spatial resolution than the Hα data. Nevertheless, the roughly constant rotation speeds of the HI at large galactocentric distances constrain the mass modeling, in particular constraining the asymptotic rotational support provided by the dark halo.

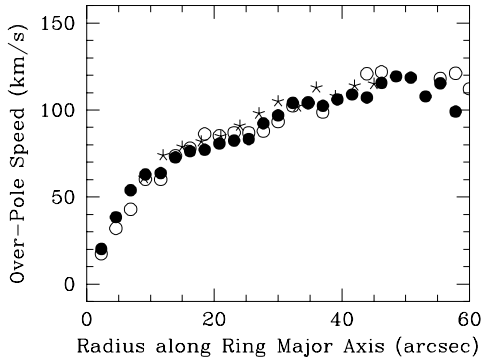

**Fig. 5.**– Over-the-pole speeds of gas orbits in the polar ring of NGC 4650A as a function of distance along the ring major axis. The data are from WMS after application of the warp correction described in §4.1. The model velocities of Nicholson (1989) are over-plotted (stars) for comparison. Approaching (*southern*) velocities are indicated by open symbols, receding (*northern*) velocities by solid symbols. The scatter in the points reflects the uncertainties.

## 3. The Model

In order to separate the contribution of the dark matter in NGC 4650A from that of the luminous matter, we construct a multi-component mass model for the galaxy consisting of a double exponential disk, a two-component (*i.e.*, stars and gas) ring, and an axisymmetric, flattened, pseudo-isothermal dark halo. Each of these components of the mass model is discussed below.

### 3.1. Thick Exponential Disk

We model the central component of NGC 4650A as a double (*i.e.*, thick) exponential disk (*cf.* Bahcall & Soneira 1980), with a mass volume density given by

$$\rho_D(R,z) = \rho_{o,D}\, e^{-R/h}\, e^{-|z|/z_o}, \tag{1}$$

where $R$ and $z$ are cylindrical coordinates in the axisymmetric disk, $h$ is the exponential scale length of the disk, and $z_o$ is the exponential scale height. The central density $\rho_{o,D}$, is then related to the total disk mass, $M_D$, through $M_D = 4\pi z_o \rho_{o,D} h^2$. The corresponding potential that solves Poisson's equation for this density distribution is (SS):

$$\Phi_D(R,z) = -GM_D \int_o^\infty \frac{dk\ J_o(kR)\left[kz_o\ e^{-|z|/z_o} - e^{-k|z|}\right]}{[1+h^2k^2]^{3/2}\ [k^2 z_o^2 - 1]} \ . \tag{2}$$

Derivatives of this potential, which we evaluate numerically, give the corresponding gravitational force for the disk at arbitrary positions in the polar and equatorial planes.

The thickness of the disk is a matter of some importance for the modeling, not only because it enters directly in the density distribution (and thus the potential) as the scale height, but also because it affects the inferred inclination and radial scale length of the disk. Thus, in order to place upper and lower limits on the axis ratio of the dark matter distribution we must constrain the axis ratio of the light distribution.

In the thin disk approximation (*i.e.*, $q \equiv z_o/h << 1$), the disk has a scale length of about $4.7''$ and an axis ratio of $q = 0.42$ derived from the axis ratio of the outer isophotes (WMS). We derive nearly identical structure parameters from our I-band image (Fig. 2). Fouqué *et al.* (1990) have argued, however, that $q_0$ is type-dependent, and suggest that for galaxies of type T, where $-5 \leq T \leq 7$, the empirical dependence is of the form $\log q_0^{-1} = 0.43 + 0.053\ T$. Since in their system, an elliptical corresponds to $T = -5$, an E/S0 to $T = -3$, and an S0 to $T = -2$, even a conservative typing of the central galaxy of NGC 4650A as an S0 would yield a thickness parameter consistent with the central object being edge-on, with a thickness equal to the measured axis ratio of the light. Therefore, we will take two separate, self-consistent and extreme choices: (1) a "thin" disk with thickness ratio $q_0 = 0.2$, and (2) a "thick" disk with $q_0 = 0.4$.

We estimate the disk luminosity and scale length using surface brightness profiles along the major and minor disk axes and extrapolating over the dusty region where the ring passes in front of the disk. By integrating Eq. 1 along a line-of-sight that is assumed to be inclined at an angle $i$ with respect to the normal of the disk, one can derive integral expressions for the projected surface brightness of a double exponential disk along its major and minor axes, and compare these to the measured surface brightness profiles. These estimates will depend on the assumed thickness and inclination of the disk.

Shown in Fig. 3 are the model predictions for the surface brightness profiles along the major and minor axes from the thin and thick disk assumptions used here. We have not done formal fits, but instead have constrained the thick and thin disk models by requiring that both models have the same total luminosity and reproduce the minor and major axis profiles reasonably well in regions that are not likely to be strongly affected by seeing or dust. Assuming an intrinsic flattening $q_0 \equiv z_0/h = 0.2$, typical for late-type spiral disks, yields a disk inclination of 68° using the classical form: $\cos^2 i = (q^2 - q_0^2)/(1 - q_0^2)$. After correcting for external extinction (using $A_B$ = 0.54, as given in the RC3) and assuming a distance of 35 Mpc (based on our Hubble constant[3] and $v_{LG} = 2625$ km s$^{-1}$), we find that the disk luminosity is then $L \approx 3.0 \times 10^9 L_{B,\odot}$, assuming no intrinsic absorption. For a thin disk, the scale length is $4.7''$, implying a scale height of $0.94''$. In order to achieve similar major and minor axis fits with a thick disk ($q_0 = 0.4$) of the same intrinsic luminosity, we require an inclination of 78° and a scale length of $4.4''$, corresponding to a scale height of $1.76''$.

The small, central blue spike in the B-band profile was interpreted by SS as a small bulge component with a mass-to-light ratio $M/L_B \leq 0.5$ (in solar units), which was determined from disk kinematics. An alternate explanation is that this central feature results from a recent burst of star formation, perhaps due to modest inflow of gas from the polar ring. This could explain its low $M/L$ and negligible effect on the dynamical modeling, as noted by SS. For these reasons, we neglect it in our analysis.

Once the disk mass-to-light ratio, $M/L_B$, which is assumed to be constant with radius, is chosen, all model parameters for the central component of NGC 4650A are fixed. To estimate $M/L_B$ for the disk, we take the maximum disk hypothesis, which places as much mass as possible in the disk consistent with rotation curve constraints. If the disk is less than maximum, the dark halo must supply more flattening to the overall potential in order to compensate for the decreased effect of the central disk. Thus, by considering only a maximum disk, we place a *conservative* limit on the flattening of the dark halo. We find, as did SS, that the polar rotation curve tightly constrains the maximum disk mass (see Fig. 6). With our warp correction for the ring, we find that $M/L_B \sim 2.0$ in solar units for the thin maximum disk, and $M/L_B \sim 1.75$ for the thick maximum disk, corresponding to disk masses of $6.0 \times 10^9 M_\odot$ and $5.25 \times 10^9 M_\odot$, respectively.

### 3.2. Two-Component Polar Ring

The mass of the polar ring of NGC 4650A is non-negligible and must be included in dynamical modeling (SS). We model the ring with two thin annuli, one for the stellar ring (and its associated molecular hydrogen complexes), and one for the more radially-extended neutral hydrogen ring. The mass of the stellar annulus is estimated from optical photometry and CO measurements; the neutral hydrogen mass is inferred from the 21-cm VLA observations of GSK. The form of the model mass distribution is given by the difference of two disks, as described below, with the scale lengths and masses of each disk chosen so that the resulting annulus will match the total mass and surface mass profile (assumed to drop to zero in the middle) deduced from observations. With the exception of the stellar and CO mass, the ring parameters are identical to those used in SS, to which we refer the reader for more detail.

---

[3]Throughout this paper we assume a Hubble constant of $H_0 = 75$ km s$^{-1}$ Mpc$^{-1}$.

The model stellar annulus is formed from the difference of two n=2 Toomre (1962) disks with scale lengths of $35''$ and $40''$. The shape of this profile, shown in Fig. 4 after correction for warping (see §4.1), agrees roughly with the distribution of light in the ring and the placement of the bi-symmetric knots. The ring is about 40% as bright in I as the central body; we use the I-band light as a rough indicator of stellar mass to estimate the total mass of stars in the ring. Adopting the "maximum disk" masses from the previous section then yields stellar ring masses of $2.4 \times 10^9 M_\odot$ for the thin disk model and $2.1 \times 10^9 M_\odot$ for the thick disk model. Using the usual conversion, the molecular hydrogen content of the inner ring can be inferred from CO measurements (Watson, Guptill & Buchholz 1994) to be $8 \times 10^8 M_\odot < M(H_2) < 16 \times 10^8 M_\odot$. The spatial distribution of the CO more closely follows that of the blue light than the HI in the ring. Adding an intermediate value of $M(H_2) = 1.2 \times 10^9 M_\odot$ to the stellar ring mass gives $3.6 \times 10^9 M_\odot$ and $3.3 \times 10^9 M_\odot$ for the combined mass of the inner ring annulus in the thin and thick disk models, respectively.

The model HI ring is formed from the difference of two $n=1$ Toomre (*i.e.*, Kuzmin) disks with scale lengths of $20''$ and $90''$ chosen to match the profile of the HI contours on the sky (see SS). After converting the HI mass of GSK to our Hubble constant and correcting for the cosmic abundance of helium, we find the ring HI mass to be $6.4 \times 10^9 M_\odot$. Note that the total mass of the ring in stars and gas is estimated to be $\sim 1 \times 10^{10} M_\odot$, somewhat smaller than the value assumed by SS, but larger than the adopted mass of the central luminous disk.

Polar orbits interior to the bulk of the ring mass are slowed by the net outward force of the ring, while ring orbits at large radii experience a net inward force and increased over-the-pole speeds. It was suggested by SS that the self gravity of the ring may cause the depression in the observed ring speeds at $\sim 25''$. It is also possible that line-of-sight effects may play a role, as ring material at larger radii crosses the ring major axis inward of $30''$ in projection.

Our mass model for the ring is coplanar and axisymmetric in the polar plane. Due to the presence of the disk and flattened halo potentials, however, the resulting ring *orbits* are oval. The effect of this deficiency in the modeling is small. A model with an oval distribution of ring mass would require more flattening in the halo component to compensate for the quadrupole moment of the ring, which would have the opposite sense as that of the central disk. Thus, by considering the ring mass to be circular, we place a conservative limit on the halo flattening.

### 3.3. Flattened Halo

Galactic potentials that are substantially flattened toward the plane of the disk may or may be axisymmetric, that is, they may or may not have an axis ratio $b/a = 1$ in the equatorial plane. The axisymmetry of the Milky Way at large radii is a subject of controversy (*cf.* Blitz & Spergel 1991, Kuijken 1991). The apparent circularity of disk galaxies supports the hypothesis that galactic potentials do not deviate strongly from axisymmetry, as do N-body simulations that include the effects of dissipation (*cf.* Katz & Gunn 1991, Dubinski 1994). Detailed modeling of the velocity field of the equatorial HI ring in the early type galaxy IC 2006 (Franx, van Gorkom & de Zeeuw 1994) indicates that its galactic potential is axisymmetric to within 5% (*i.e.*, b/a $\leq$ 0.05) with 95% confidence.

For these reasons, we have chosen a flattened, but axisymmetric, pseudo-isothermal ellipsoid to model the dark halo of NGC 4650A. The density of such a halo, given by,

$$\rho_H(R, z) = \frac{\rho_{o,H}\, r_c^2}{r_c^2 + R^2 + z^2/q^2} \quad , \tag{3}$$

is stratified on concentric, similar ellipsoids of vertical-to-radial axis ratio $q \equiv c/a$; the radial profile is chosen so as to give asymptotically flat rotation curves in the equatorial plane with circular speed $v_H$. By integrating the density distribution described by Eq. 3, the halo mass enclosed by an ellipsoid of semi-major axis $d$ and axis ratio $q$ can be shown to be

$$M_H(<d) = \frac{v_H^2\, r_c\, q\sqrt{1-q^2}}{G \arccos q} \left[\left(\frac{d}{r_c}\right) - \arctan\left(\frac{d}{r_c}\right)\right] \quad . \tag{4}$$

This model for a flattened halo is to be contrasted with an oft-used form in which the *potential*, not the density, is stratified on concentric ellipsoids of constant axis ratio (*cf.* Binney & Tremaine, p46). Our parameterization has the advantage of providing a smooth family of flattened halo models that can be extended to infinitely thin systems. Furthermore, the N-body simulations of Katz & Gunn (1991), which include the hydrodynamics required to model the gaseous component of galaxy formation, are consistent a constant axis ratio $(c/a)$ for the density of the dark matter halo at radii beyond the stellar disk (Katz, private communication 1994). The alternative model, which has isopotentials of constant axis ratio, is computationally more convenient, but produces density profiles that are strongly "pinched" along the vertical axis for large flattenings of the potential, and requires unphysical negative densities for $q_\Phi < 0.707$.

The potential corresponding to our model halo density is given by (*cf.* SS):

$$\Phi_H(R, z) = 2\pi G q\, \rho_{o,H}\, r_c^2 \int_o^{1/q} \frac{\ln\left[1 + \frac{x^2}{r_c^2}\left(\frac{R^2}{x^2(1-q^2)+1} + z^2\right)\right] dx}{x^2(1-q^2)+1} \quad . \tag{5}$$

Although we do not know a closed form expression for this potential, its derivatives with respect to $R$ and $z$ can be analytically evaluated (with some effort) to give the forces in cylindrical coordinates:

$$F_z(R, z) = \frac{-v_H^2\, z\, \gamma}{h \arctan \gamma} \left[\frac{\arctan(\gamma\mu)}{\gamma\,\mu} - \frac{\arctan(\gamma\nu)}{\gamma\,\nu}\right] \quad , \text{and} \tag{6}$$

$$F_R(R, z) = \frac{-v_H^2 R \gamma}{h \arctan \gamma} \left[ \frac{\mu^2}{\mu^2 - 1} \left( \frac{\arctan(\gamma\mu)}{\gamma\mu} - \frac{\arctan\gamma}{\gamma} \right) - \frac{\nu^2}{\nu^2 - 1} \left( \frac{\arctan(\gamma\nu)}{\gamma\nu} - \frac{\arctan\gamma}{\gamma} \right) \right] ,$$

where
$$\gamma \equiv \frac{\sqrt{1-q^2}}{q} , \qquad \nu \equiv \sqrt{\frac{2c}{b-h}} , \qquad \mu \equiv \sqrt{\frac{2c}{b+h}}$$

and
$$\begin{aligned} h &\equiv \sqrt{b^2 - 4ac} > 0 \\ a &\equiv (1-q^2) r_c^2 \\ b &\equiv z^2 + R^2 + (1-q^2) r_c^2 , \\ c &\equiv z^2 . \end{aligned}$$

We derived these expressions for the forces in cylindrical coordinates before realizing that their counterparts in ellipsoidal coordinates had already been calculated by de Zeeuw & Pfenniger (1988) in their encyclopedic paper on potential-density pairs.

Evaluating the radial force in the equatorial plane, $F_R(R,0)$, in the limit $R \to \infty$ yields an expression relating the asymptotic circular speed in the halo equatorial plane to the central density, core radius, and axis ratio of the halo:

$$v_H^2 = \frac{4\pi G \rho_{o,H} r_c^2 q \arccos q}{\sqrt{1-q^2}} . \qquad (7)$$

In constructing the combined mass model of NGC 4650A, we fix the axis ratio $q$ of the isodensity surfaces and assume that the equatorial symmetry plane of the flattened halo is aligned with the disk plane of the central galaxy of NGC 4650A. The core radius and asymptotic speed of the halo are then constrained by the observed kinematics.

## 4. Comparing Model Kinematics to the Data

For each component of the mass model, a gravitational potential satisfying Poisson's equation is constructed. Derivatives of the combined potential provide the forces at any given point, which are used to solve for the symmetric, closed orbits in the equatorial disk and polar ring planes using the numerical shooting algorithm developed by SS. A test particle is launched tangentially at a specified position along the disk (or ring) major axis and its orbit followed by integrating the equations of motion in the combined potential. The launch speed is adjusted until the orbit closes on itself; the orbit trajectory is then recorded.

The contributions to the total potential from the disk and ring are completely specified in the models (§3.1 and §3.2). Since the measured polar speeds have smaller uncertainties and extend to larger radii than the measured kinematics of the central E/S0 body, we choose a value for the halo flattening and then use the polar data to constrain the remaining free parameters of the model: the halo core radius, $r_c$, and asymptotic circular speed of the halo in the equatorial plane, $v_H$. With all the parameters fixed by the polar kinematics, models with different halo flattening will then match the equatorial *disk* rotation curve with differing degrees of success. This provides the basis for the measurement of the flattening of the dark matter distribution of NGC 4650A.

The speeds of the fully-determined closed orbits in the equatorial plane provide a measure of the radial force in the disk. Under the assumption of radial hydrostatic equilibrium, the mean azimuthal streaming velocity is related to the radial force and radial pressure (velocity dispersion) in the disk, so that we are able to calculate the streaming velocities from the Jeans equations. These streaming velocities are then integrated along the line-of-sight and convolved with the seeing disk to produce line-of-sight velocity and dispersion profiles that can be compared directly to the new disk kinematics. (The dispersion of H$\alpha$ and HI gas is typically 10 km s$^{-1}$ or less; for this reason we have ignored the $\sim$1–2% correction for radial pressure in the ring models when comparing them to the gas ring rotation curves of NGC 4650A.)

### 4.1. The Polar Rotation Curve: Ring Warp

Since the polar ring of NGC 4650A is warped, the change in inclination with ring radius must be modeled before the over-the-pole speeds are compared to the measured values. Our warp (twist) model assumes that all orbits are polar, but that their inclinations to the sky plane vary linearly with galactocentric radius along the ring major axis such that the ring is 10° from edge-on at the center and is edge-on at 30″. This linear warp correction corresponds to an inclination of 110° at a ring radius of 90″, which roughly agrees with that inferred from the axis ratio of the HI contours at this radius (GSK). Our model is based on the assumption that the bi-symmetric knots seen at $\sim$30″ (Figs. 2 and 4) are caused by orbit crowding and superposition as the ring warps through edge-on to the observer at this radius. Such a warp falls within the range of acceptable models considered by Nicholson (1989) from fits of tilted (circular) ring models to his Fabry-Perot H$\alpha$ velocity field, and differs only slightly from that assumed by SS. Since the ring is close to edge-on over most of its optical radius, the corrections to the line-of-sight speeds are less than 2% in the range 0–60″. We have applied this warp correction to the polar data plotted in Fig. 5 and all subsequent figures.

Since the ring is warped, it is unclear whether the gas seen at small ring radii should be interpreted as gas in rotation at these positions or as gas from larger orbits seen in projection. Furthermore, differential precession of this apparently slightly off-axis polar ring in a flattened potential could result in a rapidly changing inclination inward of about 10″, making the inclination correction at these small radii, and thus the true over-the-pole velocities, uncertain. Therefore, for the purposes of fitting models to the polar data, we require that the over-the-pole speeds of the model polar orbits produce a good fit to the Hα velocities only outward of 10″ and that they also match the deprojected HI speed of 105–125 km s$^{-1}$ at 90″. Since the ring may be slightly non-polar, one might worry that differential precession would negate the assumption of closed gas orbits, but since the distribution and kinematics of the polar gas is symmetric out to ∼ 90″ and the time for one orbit is substantially shorter than that for differential precession (Sparke 1986, Sparke 1991), the kinematics of the ring will be close to that of the closed polar orbits.

### 4.2. The Disk Rotation Curve: Radial Hydrostatic Equilibrium, Line-of-Sight Integration, and Seeing Convolution

Due to their radial velocity dispersion, stars in the disk are not on closed loop orbits. In order to compare the closed model orbits to the observed velocities measured along the line-of-sight in the disk, we must first calculate the difference between the local circular speed and the mean azimuthal streaming speed of the stars. We do this by applying the conditions of radial hydrostatic equilibrium.

For an axisymmetric, steady state exponential disk in which the principal axes of the velocity ellipsoid are aligned with the cylindrical coordinate system of the disk, the difference between the square of the local circular speed, $v_c^2$, and the square of the local azimuthal streaming, $\bar{v}_\phi^2$, is given by the Jeans equations as (Binney & Tremaine 1987, p. 198),

$$v_c^2 - \bar{v}_\phi^2 = \sigma_\phi^2 - \sigma_R^2 \left(1 - \frac{R}{h}\right) - \frac{\partial \sigma_R^2}{\partial \ln R} \qquad (8)$$

where $h$ is the scale length of the disk, and $\sigma_\phi$ and $\sigma_R$ are the azimuthal and radial components of the velocity dispersion.

The difference between $v_c$ and $\bar{v}_\phi$ is often called the asymmetric drift, $v_a$. In the limit that the drift is small, that is that $v_a \ll 2v_c$, the left-hand-side of Eq. 8 reduces to $2v_a v_c$, and one obtains an expression that is often used to calculate the asymmetric drift "correction" to convert observed streaming speeds to circular rotation speeds. Although the assumption that the velocity dispersion is small compared to the streaming velocity is used to justify the use of epicyclic theory to compute the relationship between $\sigma_\phi$ and $\sigma_R$, it is important to note that the derivation of Eq. 8 is applicable regardless of the relative size of $v_c$ and $\bar{v}_\phi$, and does not require that the dispersion be small.

In most stellar disks, the dispersion decreases outward, so that the mean streaming velocity is less than the local circular speed, and the difference between the two decreases rapidly with radius. For example, if one assumes that the radial velocity dispersion profile falls exponentially in radius with a scale length twice that of the light, as has been suggested for the disks of spirals (*cf.* van der Kruit & Freeman 1986, Bottema 1990), then the last term in Eq. 8 reduces to $+ (R/h)\sigma_R^2$, and the asymmetric drift decreases exponentially in magnitude with increasing radius. The dispersion in the central disk of NGC 4650A, however, appears to be nearly constant with radius (Fig. 1), so that radial pressure support is important at all radii.

In computing the asymmetric drift for NGC 4650A, we assume that the intrinsic $\sigma_\phi$ (which is the component of the velocity dispersion measured directly in this highly inclined system) falls linearly, from a central value of 66 km s$^{-1}$ to 62 km s$^{-1}$ at one scale length. After seeing convolution and integration along the line-of-sight, the predicted profile is consistent with the observed line-of-sight dispersion (see Figs. 7–10). Such a slowly declining profile is more conservative assumption than a constant dispersion profile because falling dispersion will predict larger streaming speeds for a given halo flattening, and thus require rounder halos to fit the same equatorial rotation curve. We consider then two extreme cases to bracket the behavior of $\sigma_R$, namely, that the central stellar component is (1) a hot, isotropic system with $\sigma_R = \sigma_\phi$ at all radii, and (2) sufficiently cold (supported by rotation) that $\sigma_R$ is related to $\sigma_\phi$ through the epicyclic approximation. These ability of these two alternatives to fit the data is examined in §5, where we display comparisons between model and observed kinematics.

For the assumption of isotropic dispersion, the streaming velocities $v_\phi$ can be computed directly as a function of $R$ from Eq. 8. In the epicyclic approximation, the azimuthal and radial velocity dispersions are related through

$$\frac{\sigma_\phi^2}{\sigma_R^2} = \frac{\kappa^2}{4\Omega^2} \qquad (9)$$

where $\Omega$ and $\kappa$ are the circular and epicyclic frequencies given by

$$\Omega = \frac{v_c}{R} \qquad \text{and} \qquad \kappa^2 = R\frac{d\Omega^2}{dR} + 4\Omega^2 \quad . \qquad (10)$$

Since the mass models provide the circular speed, $v_c$, as a function of $R$, we compute $\kappa$ directly from the model rotation curve and its derivatives via Eqs. 10, express $\sigma_R^2$ as a function of $\sigma_\phi^2$ through equation Eq. 9, and then compute the streaming velocities from Eq. 8.

Once the streaming velocities appropriate for each model have been computed in the plane of the central stellar component, they are projected onto the sky plane, integrated self-consistently along the line-of-sight through the thick, inclined disk, and convolved with seeing to produce the model line-of-sight velocities at each position that will be compared directly to the observations. The line-of-sight integration is performed using streaming velocities and dispersions from the modeling to construct a three-dimensional model for the luminous disk. For points above and below the equatorial plane of the disk, the galaxy is assumed to be in cylindrical rotation. The intensity at any point is assumed to be that of a thick exponential disk with the inclination, disk scale length and scale height appropriate for the "thin" and "thick" disk assumptions of §3.1. The intrinsic spectral lines are assumed to be Gaussian and isotropic. The integration is performed along a path length equivalent to 8 disk scale heights, and is convolved with a (circular) Gaussian with FWHM = 2.25″ to model seeing. The resulting line-of-sight velocities and line-of-sight dispersions are then compared directly to the observations.

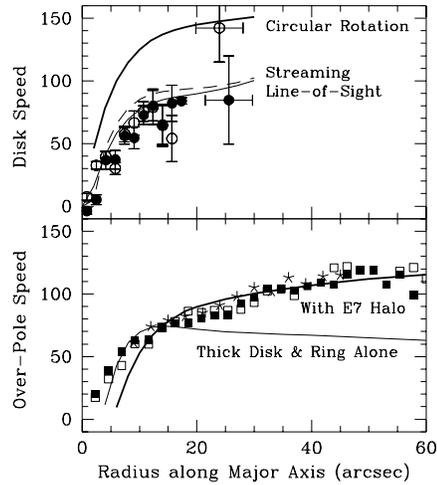

Fig. 6.– **Top Panel:** The effect of velocity corrections on the "Thick Disk + Ring + E7 Halo" model with isotropic velocity dispersion. The thick solid line shows the circular rotation speed as a function of radius for this model. The dashed line indicates the streaming velocity after radial hydrostatic equilibrium has been enforced, and the thin solid line indicates line-of-sight velocities after correction for inclination, integration along the line-of-sight, and seeing convolution. The data are the new disk kinematics presented in §2.1. **Bottom Panel:** The thick solid line shows the over-the-pole speeds for the same model superposed on the warp-corrected polar kinematics; the thin line indicates over-the-pole speeds for the thick maximum disk plus ring alone.

The effects of the pressure support (velocity dispersion), seeing convolution, and integration along the line-of-sight are illustrated in the top panel of Fig. 6. In this figure, the thick solid line indicates the circular rotation curve for the "Thick Disk + Ring + E7 Halo" model. The dashed line shows the predicted streaming speeds after radial hydrostatic equilibrium has been enforced under the assumption of an isotropic velocity ellipsoid. The thin solid line shows the line-of-sight speeds after these streaming speeds have been convolved along the line-of-sight and with the seeing disk. Since the velocity dispersion does not fall appreciably with radius, pressure support is important at all radii. For nearly edge-on systems like the disk of NGC 4650A, the $(1/\sin i)$ inclination correction to the velocities is quite small, but the convolution effects can be substantial, especially at small radii where the line of sight samples material at substantially differing radii and the seeing disk contains regions of the galaxy that are far from the major axis of the disk.

## 5. The Shape of the Dark Halo

The kinematics of NGC 4650A clearly indicate the presence of dark matter. In the bottom panel of Fig. 6, the over-the-pole speeds for a "Thick Disk + Ring + E7 Halo" model with isotropic dispersion are overplotted (thick solid line) on the warp-corrected polar ring kinematics. Also shown are the over-the-pole speeds that would be expected for the disk and ring alone (thin solid line). The gravitational force provided by the luminous matter alone is clearly insufficient to support the large rotation speeds observed in the polar ring of NGC 4650A at large galactocentric radius.

The results for our four models — the maximum "thin" and "thick" disks considered in §3.1, and the two different models for the velocity dispersion ellipsoid, namely the epicyclic and isotropic assumptions detailed in §4.2 — are presented in Figs. 7–10. The figures display, for four different halo flattenings, fits to the over-the-pole speeds in the ring (top panel), the resulting line-of-sight velocities in the disk (middle panel), and the assumed line-of-sight disk dispersions (bottom panel), superposed on the data points. Since the interpretation of the polar line-of-sight speeds is most uncertain inward of $10''$, we do not use these points to constrain the models and do not display the model curves in this radial range. As the top panels of Figs. 7–10 illustrate, by adjusting the otherwise unconstrained halo parameters $r_c$ and $v_H$, very similar fits to a single rotation curve (in this case, the polar curve) can be achieved with halos of different flattening. Another constraint (here, the equatorial disk rotation curve) is required to break this degeneracy. Flattened halos produce orbits that are elongated along the polar axis and have over-the-pole speeds that are lower than those in a spherical halo with the same $r_c$ and $v_H$. Thus, in order to reproduce the same *polar* rotation curve, flattened halos will require larger $v_H$ and smaller $r_c$.

For a given halo flattening and disk model, the combination of the well-sampled H$\alpha$ polar rotation curve from $10$–$60''$ and the polar HI speeds at large distances allow us to determine $v_H$ to typically 5 km s$^{-1}$ and $r_c$ to typically $1''$; we give $r_c$ and $v_H$ to this precision in Table 3 for each of the four halo flattenings (E5, E6, E7, and E8) considered. The central density of the halo, which depends on these two parameters and the halo axis ratio, can be computed from the relationship given by Eq. 7 in §3.3. The amount of dark matter enclosed in each model by an ellipsoid of semi-major axis $d$ and axis ratio $q$ can be computed via Eq. 4; this quantity is tabulated in Table 3 for all models at a fiducial distance of $d = 60''$ as measured in the equatorial plane. Mass outside this ellipsoid does not affect circular rotation in the equatorial plane interior to $d$. Since for distances much larger than the core radius of the halo, the mass of an oblate halo grows linearly with distance along a given axis, significantly more dark matter is required to explain the deprojected HI velocities of $\sim 120$ km s$^{-1}$ at *polar* galactocentric distances of $90$–$120''$. Note that for flattened halos, the usual measure of the total enclosed mass responsible for a circular speed $v_e$ at a distance $R$ in the equatorial plane, namely $v_e^2 R/G$, is not reliable.

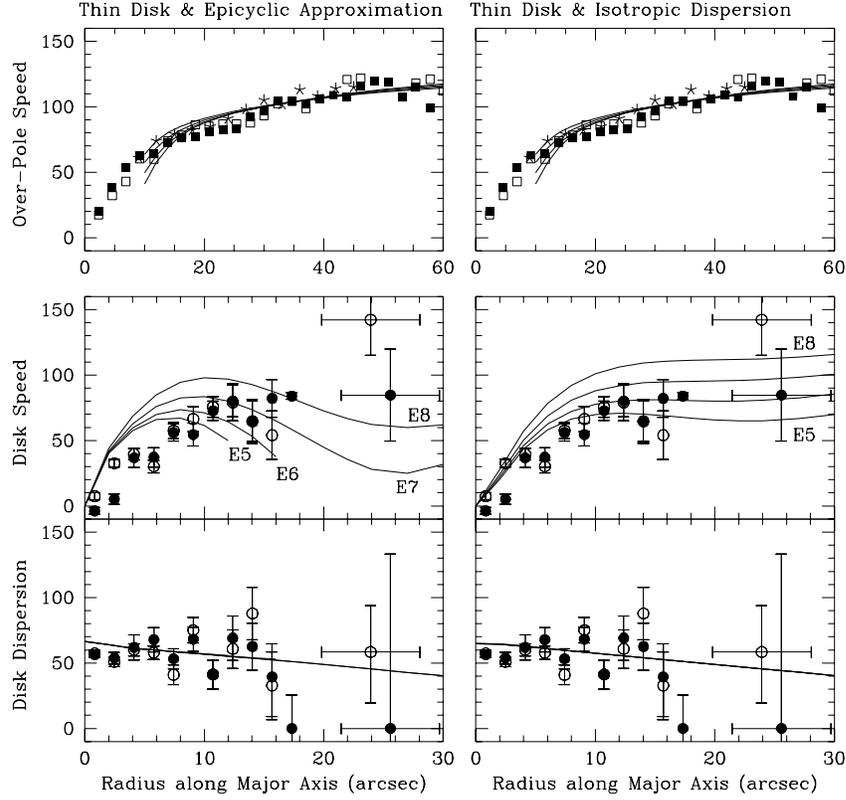

**Figs. 7. and 8.– Left (Fig. 7):** Thin Disk ($i = 68°$, $q_o = 0.2$) with a velocity ellipsoid consistent with the epicyclic approximation. **Top Panel:** Model over-the-pole speeds as a function of distance along the major axis of the ring for potentials consisting disk, ring, and E5, E6, E7, or E8 halos, superposed on the polar data of WMS (squares) and Nicholson (stars). **Middle Panel:** Model line-of-sight velocities along the major axis of the disk in the same combined potentials, superposed on the observed velocities from the new disk kinematics presented here. **Bottom Panel:** Model line-of-sight velocity dispersion in the disk superposed on the new data. Note the different spatial scales between the top and bottom two panels. **Right (Fig. 8):** Same as Fig. 7 but for the Thin Disk ($i = 68°$, $q_o = 0.2$) with isotropic velocity dispersion.

The most striking result of this work is that all models with dark halos rounder than E6 or flatter than E7 are unable to reproduce the kinematics of NGC 4650A. Within the context of these models, then, the axis ratio of the isodensity surfaces the dark halo of this galaxy are constrained to lie between $c/a = 0.3$ and $c/a = 0.4$. We find (as did SS) that the fits to the polar rotation curve improve slightly if the disk is not maximal; we chose maximal disks in order to place the largest amount of the flattening in the disk as possible, thereby enabling us to place a conservative limit on the flattening of the dark matter halo. Models with a "cold" central disk for NGC 4650A that make use of the epicyclic approximation to link the radial and azimuthal velocity dispersions (Figs. 7 and 9) are unable to reproduce the nearly linear rise of the observed disk speeds at small radii; a "hot" disk with isotropic dispersion (Figs. 8 and 10) does much better. As one would expect, a thinner disk (Figs. 7 and 8) requires a slightly rounder halo to produce the same general amplitude for the disk rotation curve, but the overall fit is better for the thick disk models (Figs. 9 and 10). Despite the smaller radial scale length $h$ of the thick disk, it is better able to reproduce the slow rise in observed disk speeds because its increased inclination increases the effect of line-of-sight and seeing convolution. The best fits to all available photometry and kinematics of NGC 4650A are thus provided by mass models that combine a massive polar ring with a thick, highly inclined disk and an E6–E7 halo (Fig. 10).

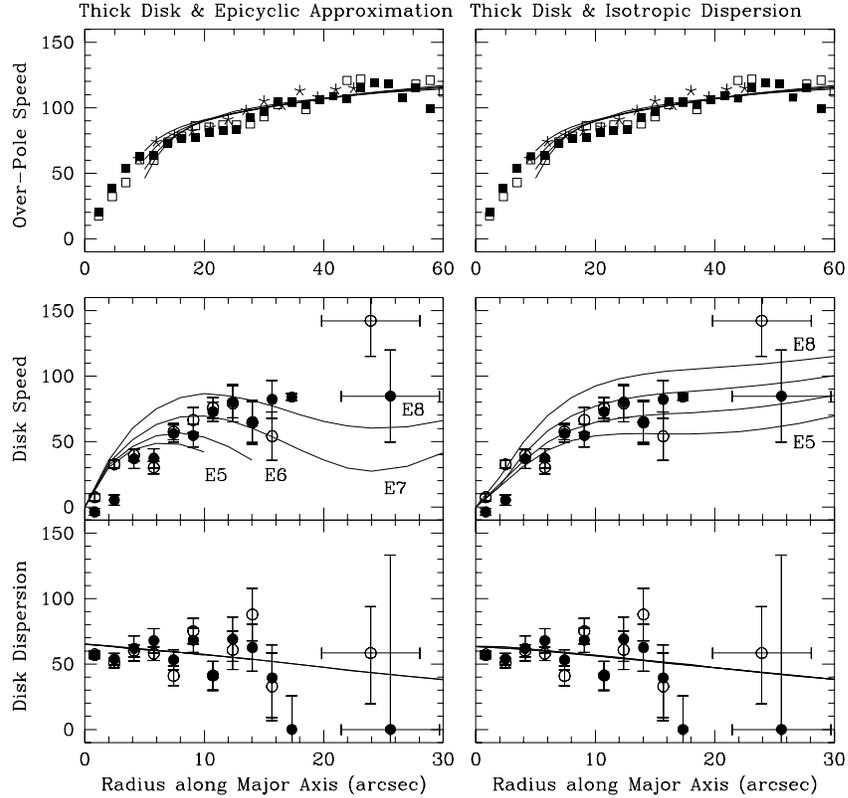

**Figs. 9. and 10.–** Left (**Fig. 9**): Same as Fig. 7 but for the Thick Disk ($i = 78°$, $q_o = 0.4$) with a velocity ellipsoid consistent with the epicyclic approximation. **Right** (**Fig. 10**): Thick Disk ($i = 78°$, $q_o = 0.4$) with isotropic velocity dispersion.

As an independent check on the flattening of the mass distribution of NGC 4650A at large galactocentric distance, we can ask whether its circular rotation speed in the equatorial plane is in agreement with the magnitude-HI linewidth (or Tully-Fisher) relation derived for disk galaxies. Although it is by no means certain that this magnitude-linewidth relation can be applied to polar-ring galaxies, it does work well over a wide variety of galaxy types. The relation is calibrated using spirals whose gas resides in the equatorial plane. The gas that defines the width of the line profile is located at large radii on the flat part of the rotation curve, and thus rotates with the asymptotic equatorial speed of the galaxy. If the galactic potential is spherical, the magnitude-linewidth relation should be independent of the plane in which the gas rotates. If the mass distribution is highly flattened, on the other hand, gas on polar orbits around a galaxy viewed edge-on would be expected to have smaller linewidth than spirals with the same optical magnitude.

Table 3: Model Parameters

| Halo Flattening | $r_c$ ($''$) | $v_H$ (km s$^{-1}$) | $v_p(120'')$ (km s$^{-1}$) | $v_e^2 R(60'')/G$ ($10^9 M_\odot$) | $M_H(<60'')$ ($10^9 M_\odot$) |
|---|---|---|---|---|---|
| \multicolumn{6}{c}{Figs. 7 and 8: Thin Disk} | | | | | |
| \multicolumn{6}{c}{$i=68°$, $M_D = 6.0 \times 10^9 M_\odot$, $h=4.7''$, $q_o = 0.2$, $M_R = 10 \times 10^9 M_\odot$} | | | | | |
| E5 | 15 | 160 | 123 | 50 | 17 |
| E6 | 13 | 165 | 122 | 56 | 14 |
| E7 | 11 | 170 | 120 | 62 | 12 |
| E8 | 9 | 175 | 118 | 68 | 8 |
| \multicolumn{6}{c}{Figs. 9 and 10: Thick Disk} | | | | | |
| \multicolumn{6}{c}{$i=78°$, $M_D = 5.25 \times 10^9 M_\odot$, $h=4.4''$, $q_o = 0.4$, $M_R = 9.7 \times 10^9 M_\odot$} | | | | | |
| E5 | 14 | 160 | 123 | 50 | 16 |
| E6 | 12 | 165 | 122 | 56 | 13 |
| E7 | 10 | 170 | 120 | 63 | 11 |
| E8 | 8 | 175 | 119 | 69 | 7 |

To apply this test to NGC 4650A, we use our measured I-band magnitude of $m_I = 11.43$ and a distance based on the redshift (corrected to the Local Group) of $v_{LG} = 2625$ km s$^{-1}$ to predict the asymptotic, equatorial speed of NGC 4650A from the I-band magnitude-linewidth relation of Guhathakurta *et al.* (1993). This value can then be compared to the equatorial speed derived from our mass models of differing halo flattening (Table 3). We note that NGC 4650A does not appear to be located in a region of strong peculiar flow, since a distance of 2725 ± 71 km s$^{-1}$ is derived from a smooth radial velocity field based on POTENT and Tully-Fisher and D-$\sigma$ distances (Bertschinger, private communication 1994). With the ring light included and no correction for internal extinction, the predicted equatorial linewidth is 366 km s$^{-1}$; this width can only be matched by halo models flatter than E8. Excluding the ring light gives 326 km s$^{-1}$ (matched by E5–E6 halo models). If the internal extinction correction used by Guhathakurta et al. (1993) is applied, the predicted linewidth is 383 km s$^{-1}$ (flatter than E8) with the ring light and 341 km s$^{-1}$ (E7) without. Using the bulk-flow-corrected distance increases the predicted linewidth by 3%. The Guhathakurta et al. relation has a 1-$\sigma$ scatter of 0.13 magnitudes, which corresponds to 16 km s$^{-1}$ in the linewidth. Note that if the potential of NGC 4650A were spherical, one would expect these estimates for the equatorial linewidth to match the measured polar HI linewidth of 240 km s$^{-1}$, which they do not: a spherical halo is inconsistent at the 5–10$\sigma$ level. Since it is likely that the ring represents delayed infall of material into a pre-existing galactic potential dominated by the dark halo, estimates that do not include the ring light may be the most appropriate. The magnitude-linewidth relation would thus predict a linewidth of 334 ±24 km s$^{-1}$ (an equatorial circular rotation speed of 167 ±12 km s$^{-1}$) for NGC 4650A, arguing strongly against a spherical dark halo, but in remarkable agreement with our result of an E6–E7 dark halo based on detailed mass modeling.

## 6. Discussion and Conclusions

Dark matter in galaxies dominates at large radii, where most rotation curves exhibit only principal two features: a turn-over radius and an asymptotic speed. This means that even if the structure parameters for the luminous matter are completely specified, a single rotation curve constrains only two of the three parameters required to define an ellipsoidal, pseudo-isothermal dark halo: the core radius, $r_c$; asymptotic equatorial rotation speed, $v_H$; and density axis ratio, $q$. The additional constraints in PRG provided by the orthogonal polar ring kinematics break this degeneracy allowing one to determine the flattening of the dark halo.

Our detailed analysis of existing data and new absorption-line disk kinematics for the polar ring galaxy NGC 4650A indicates that the dark matter halo of this system is significantly flattened toward the plane of the central body. A good fit to all available data for NGC 4650A is provided by a thick stellar exponential disk, (with $i = 78°$, $M_D = 5.3 \times 10^9\,M_\odot$, $h = 4.4''$, $z_0 = 1.76''$, and a slowly declining isotropic velocity dispersion $\sigma \sim 60$ km s$^{-1}$), surrounded by a massive, warped polar ring of gas and stars (with mass $M_R \sim 9.7 \times 10^9 M_\odot$ warping through edge-on at ring radius $\sim 30''$), and a flattened, pseudo-isothermal dark halo with density axis ratio $0.3 \lesssim c/a \lesssim 0.4$ (and $r_c = 10$–$12''$ and $v_H = 165$–$170$ km s$^{-1}$). Wherever the data were ambiguous, we have attempted to err on the side of favoring rounder halos: (1) maximum disk solutions were used throughout, (2) a slowly declining line-of-sight velocity dispersion was adopted although a flat profile is also consistent, and (3) the mass model for the polar ring was taken to be axisymmetric rather than oval. In this sense, the conclusion that the dark matter of NGC 4650A is flattened toward the central plane with an E6–E7 shape is conservative.

The central body of NGC 4650A was originally typed by Lausten & West (1980) as an E6–E7 elliptical. Since the axis ratio of the projected light from the central body is about 0.4, the intrinsic shape of the luminous body must be E6 or flatter, but as an early type galaxy (E/S0), its axis ratio is almost certainly larger than 0.3 (E7 or rounder). Although discovery of its rotation led to an S0 typing for the central body of NGC 4650A, a more correct description may be an oblate rotator with an exponential profile. Indeed, the substantial and nearly constant line-of-sight dispersion with radius might be the consequence heating during the interaction that produced the polar ring. The value of $v/\sigma \sim 1$ that we observe at the half-light radius of the central component of NGC 4650A is consistent with both an isotropic oblate rotator (see Binney & Tremaine 1987, p. 217) with an axis ratio $c/a = 0.6$ (E4) and an anisotropic rotator (isotropy parameter $\delta = 1/2$) with $c/a = 0.3$ (E7). Taken together with the main conclusion of this paper — that the dark halo of this galaxy has a flattening of E6–E7 – this leads one to the remarkable conclusion that the light and dark matter in NGC 4650A have nearly identical flattening.

The notion that the radial distributions of dark and luminous matter in galaxies are correlated so as to "conspire" to produce flat rotations was first suggested by Bahcall & Casertano (1985). The conspiracy may not be as devious or ubiquitous as it first appeared and may be partially explained by the secondary infall of halo material during disk formation (for a review, see Ashman 1992 and references therein), but dark and luminous radial structure parameters are clearly related. The coincidence of shapes for the dark and luminous matter in NGC 4650A may be an indication that the *vertical* distribution of light and dark mass is also correlated; more galaxies must be studied to ascertain whether this is a general property of galactic structure.

The shape of dark matter halos provides a clue as to the nature of the dark constituents. In recent N-body simulations of dissipationless collapse, triaxial dark halos are formed preferentially over spherical halos (*e.g.*, Frenk *et al.* 1988, Dubinski & Carlberg 1991, Warren *et al.* 1992). When a small fraction ($\sim$10%) of dissipative gas is added to these simulations, the resulting halos are more nearly oblate than prolate, and outside the inner several kiloparsec have vertical-to-radial axis ratios of $c/a \gtrsim 0.5$ (Katz & Gunn 1991, Dubinski 1994). Thus, dark halos significantly flatter than E5 may suggest that the dark matter has undergone *dissipational* collapse, which would be more natural in the context of baryonic models of dark matter. It is therefore of primary importance to determine whether the highly flattened halo of NGC 4650A is typical.

The principal results and implications of our observations and analysis of polar ring galaxy NGC 4650A can be summarized in three main points, subject to the assumptions inherent in the modeling:

• The dark matter halo of NGC 4650A, if well-represented by an pseudo-isothermal ellipsoid, has isodensity surfaces that have a vertical-to-radial axis ratio $0.3 \lesssim c/a \lesssim 0.4$, *i.e.*, has a shape between $\sim$E6 and $\sim$E7.

• The central luminous object of NGC 4650A is an early-type system with an intrinsic axis ratio that is quite similar to that of its dark matter halo.

• The flattening that we infer for the dark halo of NGC 4650A is larger than that predicted from N-body collapse simulations of dissipationless dark matter.

It is a pleasure to express gratitude to Linda Sparke for many useful comments bearing on this project and for a critical reading of the manuscript. We thank Marijn Franx, Masataka Fukugita, Raja Guhathakurta, and Brad Whitmore for valuable discussions, and Ed Bertschinger for providing the software enabling us to estimate the peculiar velocity of NGC 4650A. Observations reported in this work were carried out at the European Southern Observatory, La Silla Chile, and the Mt. Stromlo and Siding Springs Observatory, Australia. Work by PDS was supported in part by the National Science Foundation (AST 92-15485) and a grant from the J. Seward Johnson Charitable Trust. HWR was supported by a Hubble fellowship (HF-1024.01-91A).